\documentclass[aps,prd,preprint,groupedaddress]{revtex4-2}

\usepackage[utf8]{inputenc}
\usepackage{amsmath}
\usepackage{physics}
\usepackage{mathtools}
\usepackage{amssymb}
\usepackage{slashed}
\usepackage{tikz-feynman}
\usepackage{amsthm}
\usepackage{txfonts}

\begin{document}

\title{Comment on: “Geometric phase of neutrinos: differences between Dirac and Majorana neutrinos” [Phys. Lett. B \textbf{780}, 216  (2018)]}

\author{Jianlong Lu}
\email[]{jianlong\_lu@u.nus.edu}

\affiliation{Department of Physics, National University of Singapore. 2 Science Drive 3, Singapore 117551}

\date{\today}

\begin{abstract}
In [Phys. Lett. B \textbf{780}, 216  (2018)], Capolupo et al. calculate the noncyclic geometric phases associated with neutrino flavor transitions in the two-neutrino scenario, and claim that the results depend on the choice of representations of neutrino mixing matrix thus can be use to distinguish between Dirac and Majorana neutrinos. In this paper, we show that their calculation is flawed due to incorrect application of the definition of noncyclic geometric phase and the omission of one term in Wolfenstein effective Hamiltonian. Their results are neither gauge invariant nor lepton field rephasing invariant. We present an alternative calculation, which is solely in order to demonstrate that the Majorana CP-violating phase enters the geometric phase essentially by lepton field rephasing transformation. We point out that the seemingly nontrivial dependence of geometric phase on Majorana CP-violating
phase presented in [Phys. Lett. B \textbf{780}, 216 (2018)] is unphysical and thus unmeasurable.
\end{abstract}
\maketitle

\section{}
As a generalisation of the Berry phase, Mukunda and Simon defined the noncyclic geometric phase for a state vector $\ket{\psi(s)}$ evolving along a path $\Gamma$ in Hilbert space with real parameter $s\in[s_{0},s_{1}]$ as follows \cite{mukunda}:
\begin{align}
\label{eqn:def}
   \Phi^{g}_{\psi} (\Gamma)= \Phi^{t}_{\psi}(\Gamma) - \Phi^{d}_{\psi}(\Gamma) = \arg(\bra{\psi(s_{0})}\ket{\psi(s_{1})}) - \Im(\int_{s_{0}}^{s_{1}}\bra{\psi(s)}\ket{\dot{\psi}(s)}ds),
\end{align}
where $ \Phi^{t}_{\psi}(\Gamma) \equiv  \arg(\bra{\psi(s_{0})}\ket{\psi(s_{1})})$ and $\Phi^{d}_{\psi}(\Gamma)\equiv \Im(\int_{s_{0}}^{s_{1}}\bra{\psi(s)}\ket{\dot{\psi}(s)}ds)$ are called total phase and dynamical phase respectively. It has been shown that $ \Phi^{g}_{\psi} (\Gamma)$ defined above is invariant under the gauge transformation 
\begin{align}
\label{eqn:gauge}
    \ket{\psi(s)} \rightarrow \ket{\psi'(s)} = e^{i\phi(s)} \ket{\psi(s)} 
\end{align}
with any smooth real function $\phi(s)$. The key point is that, under the above transformation, the integrand of the second term in Eq. (\ref{eqn:def}) undergoes the change 
\begin{align}
   \Im(\bra{\psi(s)}\ket{\dot{\psi}(s)})\rightarrow  \Im(\bra{\psi'(s)}\ket{\dot{\psi'}(s)}) =  \Im(\bra{\psi(s)}\ket{\dot{\psi}(s)}) + \dot{\phi}(s). 
\end{align}
After integration along the path $\Gamma$, the extra term $\dot{\phi}(s)$ leads exactly to the extra phase difference between the final state and initial state caused by the gauge transformation, i.e., $ \arg(\bra{\psi'(s_{0})}\ket{\psi'(s_{1})})- \arg(\bra{\psi(s_{0})}\ket{\psi(s_{1})})$. When $\Gamma$ is a loop, i.e., $\ket{\psi(s_{0})}=\ket{\psi(s_{1})}$, the first term in Eq. (\ref{eqn:def}) vanishes and $\Phi^{g}_{\psi} (\Gamma)$ reduces to the familiar Berry phase \cite{berry}  \cite{wilczek}.\\
In the simplified scenario with only two generations of neutrinos, the general $2\times 2$ unitary mixing matrix can be parametrised as \cite{giunti1}
\begin{align}
   U= \begin{pmatrix}\cos\theta e^{i\omega_{1}} & \sin\theta e^{i(\omega_{1}+\alpha)} \\ -\sin\theta e^{i(\omega_{2}-\alpha)} & \cos\theta e^{i\omega_{2}}    \end{pmatrix}.
\end{align}         
Two unphysical degrees of freedom hidden under $(\omega_{1}, \omega_{2}, \alpha)$ can be removed by rephasing the charged lepton fields, due to the existence of charged-current weak interaction which glues together the charged lepton fields and neutrino fields. Because of the invariance of charged-current interaction Lagrangian, charged lepton field rephasing naturally implies corresponding simultaneous flavor neutrino field rephasing. Two common representations of the mixing matrix are as follows:     
\begin{align}
     U^{(1)}=\begin{pmatrix}\cos\theta & \sin\theta e^{i\alpha} \\ -\sin\theta  & \cos\theta e^{i\alpha}    \end{pmatrix},\ \ \ 
     U^{(2)}=\begin{pmatrix}\cos\theta & \sin\theta e^{i\alpha} \\ -\sin\theta e^{-i\alpha}   & \cos\theta    \end{pmatrix}.
\end{align}
These two representations are adopted in \cite{capolupo1}. If matter effect with constant matter density is significant \cite{wolfenstein} \cite{giunti2}, the mixing angle $\theta$ and the mass-squared difference $\Delta m^{2}$ are respectively modified to $\theta_{m}$ and $\Delta \tilde{m}^{2}$ due to the effect of coherent forward scattering, which satisfy \cite{XZZ}
\begin{align}
    \tan 2\theta_{m} = \frac{(m_{2}^{2}-m_{1}^{2})\sin 2\theta}{(m_{2}^{2}-m_{1}^{2})\cos 2\theta - 2\sqrt{2}G_{F}n_{e}E},
\end{align}
and 
\begin{align}
   \Delta \tilde{m}^{2} = (m_{2}^{2}-m_{1}^{2}) \sqrt{(\cos 2\theta -\frac{2\sqrt{2}G_{F} n_{e} E}{m_{2}^{2}-m_{1}^{2}})^{2}+\sin^{2} 2\theta},
\end{align}
where $n_{e}$ is the electron number density in the medium. For convenience of reference, we denote the modified mixing matrix by $U^{{\rm m}}$, with two different representations $U^{{\rm m}(1)}$ and $U^{{\rm m}(2)}$. The effective Hamiltonian can be written as 
\begin{align}
\label{eqn:effh}
    \mathcal{H}^{{\rm m}} = \Big(\frac{m_{1}^{2}+m_{2}^{2}}{4E} + \frac{G_{F}n_{e}}{\sqrt{2}} \Big) \textbf{1} + \mathcal{H}_{I}.
\end{align}
Corresponding to $U^{{\rm m}(1)}$ and $U^{{\rm m}(2)}$, there are two representations of $\mathcal{H}_{I}$, denoted by $\mathcal{H}_{I}^{(1)}$ and $\mathcal{H}_{I}^{(2)}$ respectively. One has 
\begin{align}
       \mathcal{H}_{I}^{(1)} = \begin{pmatrix} \frac{G_{F}n_{e}}{\sqrt{2}} - \frac{\Delta m^{2}}{4E}\cos 2\theta & \frac{\Delta m^{2}}{4E}\sin 2\theta  \\  \frac{\Delta m^{2}}{4E}\sin 2\theta & -\frac{G_{F}n_{e}}{\sqrt{2}} +\frac{\Delta m^{2}}{4E}\cos 2\theta   \end{pmatrix},\ \
       \mathcal{H}_{I}^{(2)} = \begin{pmatrix} \frac{G_{F}n_{e}}{\sqrt{2}} - \frac{\Delta m^{2}}{4E}\cos 2\theta & \frac{\Delta m^{2}}{4E}\sin 2\theta e^{i\alpha}\\  \frac{\Delta m^{2}}{4E}\sin 2\theta e^{-i\alpha} & -\frac{G_{F}n_{e}}{\sqrt{2}} +\frac{\Delta m^{2}}{4E}\cos 2\theta   \end{pmatrix}.
\end{align}
It is easy to verify that $\mathcal{H}_{I}^{(1)}$ and $\mathcal{H}_{I}^{(2)}$ are diagonalised respectively by $U^{(1)}$ and $U^{(2)}$, with the same spectrum, i.e.,
\begin{align}
     (U^{{\rm m}(1)})^{\dagger} \mathcal{H}_{I}^{(1)} U^{{\rm m}(1)} = (U^{{\rm m}(2)})^{\dagger} \mathcal{H}_{I}^{(2)} U^{{\rm m}(2)}= \begin{pmatrix} -\frac{\Delta \tilde{m}^{2}}{4E} & 0 \\ 0 & +\frac{\Delta \tilde{m}^{2}}{4E} \end{pmatrix}.
\end{align}
The first term of $\mathcal{H}^{{\rm m}}$ in Eq. (\ref{eqn:effh}) is proportional to the identity matrix, and thus it is irrelevant to the neutrino transition probabilities \cite{XZZ}. However, this term will lead to nontrivial contribution in the calculation of geometric phase, as shown in the next section, and thus cannot be neglected.\\
With plane wave assumption and relativistic approximation, the evolution of neutrinos initially at flavor eigenstates can be described by      
\begin{align}
\label{eqn:evo}
   \begin{pmatrix} \ket{\nu_{e}(x)} \\ \ket{\nu_{\mu}(x)}\end{pmatrix}  = (U^{{\rm m}})^{*} e^{-i(\frac{m_{1}^{2}+m_{2}^{2}}{4E} + \frac{G_{F}n_{e}}{\sqrt{2}})x}  \begin{pmatrix}   e^{+i\frac{  \Delta \tilde{m}^{2} }{4E}x}\ket{\nu_{1}}\\  e^{-i\frac{  \Delta \tilde{m}^{2} }{4E}x}\ket{\nu_{2}} \end{pmatrix}.
\end{align}
Note that the complex conjugate of the mixing matrix is used in the above equation, since we are dealing with neutrino state vectors instead of neutrino fields now \cite{XZZ}.\\
This letter concerns the incorrect way of applying Eq. (\ref{eqn:def}) to neutrino flavor transitions and the unphysical origin of the dependence of geometric phase on Majorana CP-violating phase presented in \cite{capolupo1}.

\section{}
In \cite{capolupo1}, Capolupo et al. claim that the noncyclic geometric phases associated with the transitions from $\nu_{e}$ to $\nu_{\mu}$ and from $\nu_{\mu}$ to $\nu_{e}$ after spatial propagation from $0$ to $z$ are 
\begin{align}
\label{eqn:wrong1}
   \Phi^{g}_{\nu_{e}\rightarrow \nu_{\mu}} (z) = \arg(\bra{\nu_{e}(0)}\ket{\nu_{\mu}(z)}) - \Im(\int_{0}^{z}\bra{\nu_{e}(z')}\ket{\dot{\nu_{\mu}}(z')}dz'),
\end{align}
\begin{align}
\label{eqn:wrong2}
   \Phi^{g}_{\nu_{\mu}\rightarrow \nu_{e}} (z) = \arg(\bra{\nu_{\mu}(0)}\ket{\nu_{e}(z)}) - \Im(\int_{0}^{z}\bra{\nu_{\mu}(z')}\ket{\dot{\nu_{e}}(z')}dz').
\end{align}
The contribution from the first term of effective Hamiltonian $\mathcal{H}^{{\rm m}}$ is also explicitly neglected in \cite{capolupo1}. We cannot agree with their procedure. Without loss of generality, one can focus on the transition $\nu_{e}\rightarrow \nu_{\mu}$ as an example. From Eq. (\ref{eqn:evo}), it is easy to see that the two neutrino flavor eigenstates are $\ket{\nu_{e}(0)}$ and $\ket{\nu_{\mu}(0)}$. While applying Eq. (\ref{eqn:def}) to the transition $\nu_{e}\rightarrow \nu_{\mu}$, the first term should be the phase of the inner product of the initial state and the final state, which are $\ket{\nu_{e}(0)}$ and $\ket{\nu_{\mu}(0)}$ respectively; the integrand in the second term should be the inner product of the state vector and its own derivative with respect to the path parameter.  Besides the formal incorrectness, Eq. (\ref{eqn:wrong1}) and Eq. (\ref{eqn:wrong2}) are also not invariant under the gauge transformation Eq. (\ref{eqn:gauge}), which is obviously a severe problem if the authors of \cite{capolupo1} wish to utilise the geometric phases as physical observables to distinguish between neutrino natures. In the following we suggest an alternative way of calculation. We would like to emphasize that this alternative calculation is not intended to replace the one in \cite{capolupo1}, since the obtained results are not rephasing invariant and thus not experimentally relevant, as proved in the later part of this manuscript. This alternative calculation is proposed solely in order to demonstrate that the Majorana CP-violating phase enters the geometric phase essentially by unphysical rephasing transformation.\\
The flavor transition process $\nu_{e}\rightarrow \nu_{\mu}$ consists of two successive steps:\\
(1) The initial state $\ket{\nu_{e}(0)}$ evolves into the state $\ket{\nu_{e}(z-\tau)}$ after spatial propagation from $0$ to $(z-\tau)$, where $\tau$ is some small positive real number.\\ 
(2) The state $\ket{\nu_{e}(z-\tau)}$ collapses to the neutrino flavor eigenstate $\ket{\nu_{\mu}(0)}$, i.e., the muon neutrino, after measurement performed at the detector.\\
The exact process of wave function collapse is still far from clear at the present stage. We prefer not to dive into such a bottomless ocean. As a simple approximation, we model the evolution of state vector from $(z-\tau)$ to $z$ by a homogeneous process $\frac{s}{\tau}\ket{\nu_{\mu}(0)}+\frac{\tau -s}{\tau} \ket{\nu_{e}(z-\tau)}$ with $s$ running from $0$ to $\tau$. We introduce $\tau$ in order to avoid the unwanted infinity coming from $\ket{\dot{\nu_{e}}(z')}$ at the final moment due to sudden change. This approximation seems problematic when $\tau\neq 0$, since the evolution is generally not unitary. However, such non-unitarity can be safely ignored if the limit $\tau\rightarrow 0$ is taken at the end of calculation. In this way, the geometric phase associated with $\nu_{e}\rightarrow \nu_{\mu}$ is
\begin{align}
\label{eqn:etomu}
     \Phi^{g}_{\nu_{e}\rightarrow \nu_{\mu}} (0,z)&=\lim_{\tau\rightarrow 0}\Big\{ \arg(\bra{\nu_{e}(0)}\ket{\nu_{\mu}(0)}) - \Im(\int_{0}^{z-\tau}\bra{\nu_{e}(y)}\ket{\dot{\nu}_{e}(y)}dy)\nonumber\\
    &\ \ \ \ -\Im(\frac{1}{\tau^{2}}\int_{0}^{\tau}\Big[\bra{\nu_{\mu}(0)}s\ e^{-i\Omega_{e\mu}(z)}+\bra{\nu_{e}(z-\tau)}(\tau -s)\Big]\Big[e^{i\Omega_{e\mu}(z)}\ket{\nu_{\mu}(0)}-\ket{\nu_{e}(z-\tau)}\Big]ds    )\Big\},
\end{align}
in which $\Omega_{e\mu}(z)$ is the relative phase between the $\ket{\nu_{\mu}(0)}$ part and the $\ket{\nu_{e}(z-\tau)}$ part. It is easy to see that this relative phase should be some function of $z$ (and equivalently of time, since we have applied the relativistic approximation). The exact form of this function depends on the dynamics of collapse. Note that the functional form of the relative phase should generally be different for different flavor transition processes, for which we use $\Omega_{e\mu}(z)$ and $\Omega_{\mu e}(z)$ to distinguish between two different flavor transition processes. One may notice that the first term, i.e., the argument of $\bra{\nu_{e}(0)}\ket{\nu_{\mu}(0)}$, is undefined because $\ket{\nu_{e}(0)}$ and $\ket{\nu_{\mu}(0)}$ are orthogonal to each other. Fortunately, it is the phase difference instead of the individual phase value that matters in interference, if it is experimentally feasible. We can therefore safely and conveniently assume that the argument of $\bra{\nu_{e}(0)}\ket{\nu_{\mu}(0)}$ vanishes. Another point deserving attention is that in Eq. (\ref{eqn:etomu}) the phase of the final state $\ket{\nu_{\mu}(0)}$ has been prescribed. It is straightforward to restore the nontrivial phase factor by the replacement $\ket{\nu_{\mu}(0)}\rightarrow e^{i\varphi} \ket{\nu_{\mu}(0)}$ and $\bra{\nu_{\mu}(0)}\rightarrow e^{-i\varphi} \bra{\nu_{\mu}(0)}$. This is exactly the rephasing transformation that will be explicitly discussed in the later part of this manuscript. Rephasing transformation is the bridge between the two representations of neutrino mixing matrix $U^{(1)}$ and $U^{(2)}$, and also the origin of the appearance of the Majorana CP-violating phase in the geometric phase, as will be shown later. No generality is lost. Thus, for the sake of simplicity, we can safely assume $\varphi =0$ temporarily and proceed.\\
By substituting $U^{(1)}$ and $U^{(2)}$ respectively (with mixing angle modified) into Eq. (\ref{eqn:etomu}), one has
\begin{align}
    \Phi^{g}_{\nu_{e}\rightarrow \nu_{\mu}} (0,z;U^{(1)})&=  (\frac{m_{1}^{2}+m_{2}^{2}}{4E} + \frac{G_{F}n_{e}}{\sqrt{2}})z  -\frac{  \Delta \tilde{m}^{2} }{4E}z \cos 2\theta_{m}\nonumber\\
    &\ \ \ -\cos((\frac{m_{1}^{2}+m_{2}^{2}}{4E} + \frac{G_{F}n_{e}}{\sqrt{2}})z+\Omega_{e\mu}(z))\sin(\frac{\Delta\tilde{m}^{2}}{4E}z) \sin 2\theta_{m} ,
\end{align}
\begin{align}
\label{eqn:correct1}
     \Phi^{g}_{\nu_{e}\rightarrow \nu_{\mu}} (0,z;U^{(2)})&=  (\frac{m_{1}^{2}+m_{2}^{2}}{4E} + \frac{G_{F}n_{e}}{\sqrt{2}})z  -\frac{  \Delta \tilde{m}^{2} }{4E}z \cos 2\theta_{m}\nonumber\\
    &\ \ \ -\cos((\frac{m_{1}^{2}+m_{2}^{2}}{4E} + \frac{G_{F}n_{e}}{\sqrt{2}})z+\Omega_{e\mu}(z)+\alpha)\sin(\frac{\Delta\tilde{m}^{2}}{4E}z) \sin 2\theta_{m}.
\end{align}
Similarly, for the transition $\nu_{\mu}\rightarrow \nu_{e}$, it is straightforward to obtain
\begin{align}
    \Phi^{g}_{\nu_{\mu}\rightarrow \nu_{e}} (0,z;U^{(1)})&=(\frac{m_{1}^{2}+m_{2}^{2}}{4E} + \frac{G_{F}n_{e}}{\sqrt{2}})z  +\frac{  \Delta \tilde{m}^{2} }{4E}z \cos 2\theta_{m}\nonumber\\
    &\ \ \ -\cos((\frac{m_{1}^{2}+m_{2}^{2}}{4E} + \frac{G_{F}n_{e}}{\sqrt{2}})z + \Omega_{\mu e}(z))\sin(\frac{\Delta\tilde{m}^{2}}{4E}z) \sin 2\theta_{m} ,
\end{align}
\begin{align}
\label{eqn:correct2}
    \Phi^{g}_{\nu_{\mu}\rightarrow \nu_{e}} (0,z;U^{(2)})&= (\frac{m_{1}^{2}+m_{2}^{2}}{4E} + \frac{G_{F}n_{e}}{\sqrt{2}})z  +\frac{  \Delta \tilde{m}^{2} }{4E}z \cos 2\theta_{m}\nonumber\\
    &\ \ \ -\cos((\frac{m_{1}^{2}+m_{2}^{2}}{4E} + \frac{G_{F}n_{e}}{\sqrt{2}})z+ \Omega_{\mu e}(z)-\alpha)\sin(\frac{\Delta\tilde{m}^{2}}{4E}z) \sin 2\theta_{m}.
\end{align}    
An immediate observation is $ \Phi^{g}_{\nu_{e}\rightarrow \nu_{\mu}} (0,z;U^{(i)})\neq  -\Phi^{g}_{\nu_{\mu}\rightarrow \nu_{e}} (0,z;U^{(i)})$ for $i=1,2$, even if the CP-violating phase vanishes. At the first sight, this seems to contradict Eq. (\ref{eqn:def}), which tells us that two processes along the same path but with opposite directions should accumulate opposite geometric phases. However, there is no real contradiction here. The two flavor transition processes $\nu_{e}\rightarrow \nu_{\mu}$ and $\nu_{\mu}\rightarrow \nu_{e}$ are not genuinely reversed processes with respect to each other, due to the existence of measurement and wave function collapse.  \\
The authors of \cite{capolupo1} claim that, when $U^{(2)}$ is applied, the inequality of the geometric phases associated with $\nu_{e}\rightarrow \nu_{\mu}$ and $\nu_{\mu}\rightarrow \nu_{e}$ is due to the present of CP-violating phase $\alpha$, revealing CP violation. This opinion is also not correct. From Eq. (\ref{eqn:correct1}) and Eq. (\ref{eqn:correct2}), one can see that the two geometric phases are still not the same even if $\alpha=0$. As mentioned above, two processes, even if they are reversed versions with respect to each other, will generally accumulate different geometric phases. Such difference is not relevant to CP asymmetry of T asymmetry.\\
Next it is interesting to understand the origin of the dependence of the obtained geometric phase on Majorana CP-violating phase, associated with neutrino flavor transitions, when $U^{(2)}$ is chosen. Here we would like to show that its origin is the unphysical rephasing transformation. Consider the following transformation on $U^{(1)}$:
\begin{align}
\label{eqn:rephase}
    U^{(1)}\rightarrow U_{\beta}= \begin{pmatrix} 1 & 0\\ 0 & e^{-i\beta}\end{pmatrix} U^{(1)}= \begin{pmatrix}\cos\theta & \sin\theta e^{i\alpha} \\ -\sin\theta e^{-i\beta} & \cos\theta e^{i(\alpha-\beta)}    \end{pmatrix},
\end{align}
which comes from the redefinition of the phase of charged lepton field (or equivalently rephasing flavor neutrino field). According to Eq. (\ref{eqn:etomu}), the corresponding geometric phases associated with flavor transitions are
\begin{align}
\label{eqn:gene1}
      \Phi^{g}_{\nu_{e}\rightarrow \nu_{\mu}} (0,z;U_{\beta})&=  (\frac{m_{1}^{2}+m_{2}^{2}}{4E} + \frac{G_{F}n_{e}}{\sqrt{2}})z  -\frac{  \Delta \tilde{m}^{2} }{4E}z \cos 2\theta_{m}\nonumber\\
    &\ \ \ -\cos((\frac{m_{1}^{2}+m_{2}^{2}}{4E} + \frac{G_{F}n_{e}}{\sqrt{2}})z+\Omega_{e\mu}(z)+\beta)\sin(\frac{\Delta\tilde{m}^{2}}{4E}z) \sin 2\theta_{m} ,
\end{align}
 \begin{align}
 \label{eqn:gene2}
      \Phi^{g}_{\nu_{\mu}\rightarrow \nu_{e}} (0,z;U_{\beta})&=  (\frac{m_{1}^{2}+m_{2}^{2}}{4E} + \frac{G_{F}n_{e}}{\sqrt{2}})z  +\frac{  \Delta \tilde{m}^{2} }{4E}z \cos 2\theta_{m}\nonumber\\
    &\ \ \ -\cos((\frac{m_{1}^{2}+m_{2}^{2}}{4E} + \frac{G_{F}n_{e}}{\sqrt{2}})z+ \Omega_{\mu e}(z)-\beta)\sin(\frac{\Delta\tilde{m}^{2}}{4E}z) \sin 2\theta_{m} .
\end{align}
At this point, we should remind ourselves that $\beta$ is a free parameter that we can freely adjust without causing any physical effects, since $\beta$ reflects the degree of freedom from the unphysical rephasing transformation. This is analogous to the choice of point of zero potential. By comparing Eq. (\ref{eqn:gene1}) and Eq. (\ref{eqn:gene2}) with Eq. (\ref{eqn:correct1}) and Eq. (\ref{eqn:correct2}), one can immediately find that the intriguing factor $\alpha$ in Eq. (\ref{eqn:correct1}) and Eq. (\ref{eqn:correct2}) ultimately originates from the transformation Eq. (\ref{eqn:rephase}). The appearance of the Majorana CP-violating phase $\alpha$ is just a trivial consequence when we choose the special case $\beta = \alpha$ for the free parameter $\beta$. When the free unphysical parameter $\beta$ is chosen to be zero, the representation of neutrino mixing matrix is transformed in such a way that the Majorana CP-violating phase $\alpha$ drops out of the formulas entirely. Therefore, the dependence of  $\Phi^{g}_{\nu_{e}\rightarrow \nu_{\mu}} (0,z;U^{(2)})$ and $\Phi^{g}_{\nu_{\mu}\rightarrow \nu_{e}} (0,z;U^{(2)})$ on $\alpha$ is unphysical and unmeasurable. It is thus impossible to use the geometric phases associated with neutrino flavor transitions, in vacuum and matter with constant density, to distinguish between Dirac and Majorana neutrinos. This is indeed consistent with the fact that Majorana CP-violating phase is irrelevant to normal neutrino-neutrino and antineutrino-antineutrino oscillations \cite{giunti1} \cite{xzz2}. Note that the above-mentioned procedure is not the only possible one, because it involves an ad hoc approximation of measurement and wave function collapse. The advantage of this procedure is that only trivially simple calculation is needed to reveal the origin of the Majorana CP-violating phase, as one can see while carrying out the integration in Eq. (\ref{eqn:etomu}). This makes it suffice for our purpose.

\section{}
In conclusion, we have shown the incorrectness in the calculation of noncyclic geometric phases associated with neutrino flavor transitions presented in \cite{capolupo1}, with an alternative calculation provided for demonstrating that the Majorana CP-violating phase enters the geometric phase essentially by field rephasing transformation. We have pointed out the unphysical nature of the seemingly nontrivial dependence of the geometric phases on Majorana CP-violating phase claimed by Capolupo et al. in \cite{capolupo1}. For normal neutrino-neutrino and antineutrino-antineutrino oscillations in vacuum and matter with constant density, neither geometric phases nor flavor transition probabilities can be used to distinguish between Dirac and Majorana neutrinos.

\end{document}